\documentclass[a4paper,12pt]{amsart}
\usepackage{amssymb,amsfonts,amscd, amsthm}
\title[Extention field theories]{Extended cohomological field theories and\\
noncommutative Frobenius manifolds}
\author[S.M.~Natanzon]{S.M.~Natanzon}

\begin{document}

\newtheorem{theorem}{Theorem}[section]
\newtheorem*{definition}{Definition}
\newtheorem{lemma}{Lemma}[section]
\newtheorem{proposition}{Proposition}[section]
\newtheorem{corollary}{Corollary}[section]
\maketitle


\newcommand{\st}{\rightarrow}
\newcommand{\Par}{\partial}
\newcommand{\T}{\theta}
\newcommand{\A}{\alpha}
\newcommand{\B}{\beta}
\newcommand{\GG}{\Gamma}
\newcommand{\G}{\gamma}
\newcommand{\E}{\epsilon}
\newcommand{\OO}{\Omega}
\newcommand{\V}{\varphi}
\newcommand{\N}{\eta}
\newcommand{\D}{\delta}
\newcommand{\X}{\xi}
\newcommand{\F}{\psi}
\newcommand{\SI}{\sigma}
\newcommand{\DD}{\Delta}
\newcommand{\FF}{\Psi}
\newcommand{\PP}{\Phi}
\newcommand{\SQ}{\sqrt}
\newcommand{\W}{\widetilde}
\newcommand{\WW}{\Tilde}

\centerline{Independent University of Moscow, Moscow, Russia}
\centerline{Moscow State University, Moscow, Russia}
\centerline{Institute Theoretical and Experimental Physics, Moscow, Russia}

\begin{abstract}
We construct some extension ({\it Stable Field Theory}) of
Cohomological Field Theory. The Stable Field Theory is a system of
homomorphisms to some vector spaces generated by spheres and disks
with punctures. It is described by a formal tensor series,
satisfying to some system of "differential equations". In points of
convergence the tensor series generate special noncommutative
analogues of Frobenius algebras, describing 'Open-Closed'
Topological Field Theories.

{\sl MSC:} 81T, 18F

{\sl Subj. Clac.} field theory

{\sl Keywords:} Field theory; Frobenius manifold

\end{abstract}
\vspace{1cm}
\centerline{INTRODUCTION}
\vspace{0.4cm}

Cohomological Field Theories were proposed by Kontsevich and Manin
\cite{Kon-Man} for description of Gromov-Witten Classes. They proved
that Cohomological Field Theory is equivalent to Formal Frobenius
manifold. Formal Frobenius manifold is defined by a formal series
$F$, satisfying the associative equations \cite {Dij, Wit2}. In
points of convergence the series $F$ defines Frobenius algebras. The
set of these points forms a Frobenius manifold as regards to
algebras investigated (with some additional condition) by Dubrovin
\cite{Dub}.

Cohomological Field Theory is a system of special homomorphisms to
spaces of cohomology of Deline--Mumford compactifications for moduli
spaces of complex rational curves (Riemann spheres) with punctures.
In this paper it is constructed some extension of Cohomological
Field Theory. This extension ({\it Stable Field Theory}) is a system
of homomorphisms to some vector spaces generated by disks with
punctures. I conjecture that they describe relative Gromov--Witten
classes.

A Stable Field Theory is equivalent to some analog of a Formal
Frobenius manifold. This analogue is defined by formal tensor series
({\it Structure Series}), satisfying some system of "differential
equations" (including the associativity equation). In points of
convergence the Structure Series define {\it Extended Frobenius
algebras}. They are special noncommutative analogue of Frobenius
algebras. Extended Frobenius algebras describe 'Open-Closed'
Topological Field Theories \cite{Laz} of genus 0 in the same way as
Frobenius algebras describe Atiyah--Witten $2D$ Topological Field
Theories \cite{At, Wit}. Thus, Structure Series are noncommutable
analogues of Formal Frobenius manifolds.

In sections 1 and 2 it is proposed a general axiomatic of
Topological Field Theories over functors. This class involves $2D$
Atiyah--Witten \cite{At, Wit}, 'open-closed' \cite{Laz}, Klein
Topological Field Theories \cite{Al-Nat} and Cohomological Field
Theories \cite{Kon-Man, Man}.

In section 3 it is constructed and investigated a {\it Stabilising
functor} on a category of spheres with punctures, disks with
punctures and its disconnected unions. For spheres this construction
is liked to modular graphs \cite{Man} and describes a Cohomological
functor for complex rational curves. The Stable Field Theories are
defined as Topological Field Theories over Stabilizing functors.

In section 4 and 5 we prove that Stable Field Theories are in
one-to-one correspondence to {\it Systems of Disk Correlation
Functions}. Its "generating functions"  are the Structure Series

In section 6 and 7 it is demonstrated that Structure Series
generate Extended Frobenius algebras in their convergence points
and it is constructed some examples.

A part of this paper was written during the author stays at Erwin
Schr\"odinger Institute in Vena, Max-Plank-Institut in Bonn and UNED
in Madrid. I thank these organisations for support and hospitality.
I would like to thank B.Dubrovin and S.Novikov for useful
discussions of results. This research is partially supported by
grants RFBR-02-01-22004a, NSh-1972.2003.1.

\section{$2D$ categories}

\subsection {}
In this paper {\it a surface} denotes a compact surface
with or without boundary. Its connected boundary component
is called \emph{a boundary contour} of the surface.
Any orientable connected surface is homeomorphic to a
sphere with $g$ handles and $s$ holes. Such surface is
called {\it a surface of type $(g,s,1)$}. Any nonorientable
connected surface is homeomorphic to either a projective
plane with $a$ handles and $s$ holes or a Klein
bottle with $a$ handles and $s$ holes. Such surface is called
{\it a surface of type $(g,s,0)$}, where $g=a+\frac 12$ in the
first case and $g=a+1$ in the second case.

A surface with a finite number of marked points
is called {\it a stratified sur\-face}.  Marked points are
also called {\it special points} of the stratified sur\-face.
Two stratified surfaces are called {\it isomorphic}, if
there exists a homeomor\-phism of the surfaces, generating
a bijection between their special points.

By $|S|$ we denote the cardinality of a finite set $S$.  Let
$\OO$ be a surface of the type $(g, s, \E)$ and $\omega_1,...,\omega_s$
be its  boundary contours. Let $S\subset \OO$ be a finite set
of marked points and $m=|S\cap(\OO\setminus \partial\OO)|$,
$m_i=|S\cap \omega_i|$. Then the collection $G=(g, \E, m, m_1,...,m_s)$
is called {\it a type} of connected stratified surface
$(\OO, S)$.
A connected stratified surface of type $G$ is called {\it
trivial}, if $\mu=2g+m+s+\frac 12\sum m_i-2\leqslant 0$. It is
easy to prove the following statement.

{\bf Lemma 1.1.} {\sl Any trivial stratified surface is
isomorphic to a stratified surface from the list:

(1) a sphere $S^2$ without special points ($\mu=-2$);

(2) a projective plane $\mathbb R P^2$ without special points ($\mu=-1$);

(3) a disk $D^2$ without special points ($\mu=-1$);

(4) a sphere $(S^2,p)$ with a single interior special point $p$ ($\mu=-1$);

(5) a disk $(D^2,q)$ with a single special boundary point and
without special interior points ($\mu=-\frac 12$);

(6) a sphere  $(S^2,p_1,p_2)$ with two interior special points ($\mu=0$);

(7) a projective plane $(\mathbb R P^2,p)$ with a single interior special
point ($\mu=0$);

(8) a torus $T^2$ without  special points ($\mu=0$);

(9) a Klein bottle $Kl$ without special points ($\mu=0$);

(10) a disk $(D^2,p)$ with a single interior special point
and without boundary special points ($\mu=0$);

(11) a disk $(D^2,q_1,q_2)$ with two boundary  special points and
without interior special points ($\mu=0$);

(12) a M\"obius band $Mb$ without special points ($\mu=0$);

(13) a cylinder $Cyl$ without special points ($\mu=0$).}

Let $\OO$ be a stratified surface. A generic not self-intersecting
curve $\G\subset\OO$ is called {\it a cut}. Generic means that $\G$
has no special points and either is a (closed) contour without
boundary points of $\Omega$ or it is a segment, whose ends belong to
the boundary of $\OO$ and all interior points are interior points of
the surface. The  cuts form 9 topological classes described in
\cite{Al-Nat}. A set of pairwise nonintersecting cuts is called {\it
a cut system}.

Let $\GG$ be a cut system of a stratified surface $\OO$. Consider
compactification $\widetilde \OO$ of $\OO\setminus \GG$
by pairs $(x,c)$, where $x\in\G\subset\GG$ and $c$ is
a coorientation of the cut $\G$ in a neighbourhood of $x$.
Denote by $\OO/ \GG$ a surface obtained by contracting
each connected component $C_i$ of
$\widetilde \OO\setminus (\OO\setminus\GG)$ into a point
$c_i$. We assume that $\OO_\#=\OO/\G$ is
a stratified surface. Its special points are the special
points of $\OO$ and the points $c_i$.

\subsection{}
In this subsection, following \cite{Al-Nat}, we define a tensor
category $\mathcal C$ of stratified surfaces with a set
$\mathcal O$ of local orientations of special points.
{\it A set of local orientations} means that for
any special point $r\in Q$ of stratified surface $\OO$
we fix an orientation $o_r$ of its small neighbourhood.

A set $\mathcal O$ of local orientations is said {\it admissible},
iff either $\OO$ is orientable surface and all local orientations
are induced by an orientation of $\OO$ or $\OO$ is non--orientable
surface and all local orientations at all special points from any
boundary contour $\omega_i$ are compatible with one of the
orientations of $\omega_i$. Moreover, we consider that any boundary
contour contain at least one special point.

{\bf Lemma 1.2} {\sl Let $(\Omega',\mathcal O')$ and
$(\Omega'',\mathcal O'')$ be two pairs, consisting of
stratified surfaces and admissible sets of local
orientations at special points. If stratified surfaces
$\Omega'$ and $\Omega''$ are isomorphic then there exists an
isomorphism $\phi:\Omega'\to\Omega''$
such that $\phi(\mathcal O')=\mathcal O''$.}

The proof follows from standard properties of surfaces.

Pairs $(\Omega,\mathcal O)$ of stratified surfaces $\OO$ with sets of
local orientations $\mathcal O$ at their special points
are {\it objects of the basic category} $C$.
{\it Morphisms} are any combinations of the morphisms of types 1)-4).

1) {\it Isomorphism} $\phi:(\Omega,\mathcal O)\to(\Omega',\mathcal O')$.
By a definition, $\phi$ is an isomorphism $\phi:\Omega\to\Omega'$ of
stratified surfaces compatible with local orientations at special points.

2) {\it Change of local orientations}
$\psi:(\Omega,\mathcal O)\to(\Omega,\mathcal O')$.
Thus, there is one such morphism for
any pair $(\mathcal O,\mathcal O')$ of sets of local orientations
on a stratified surface $\Omega$.

3) {\it Cutting} $\eta:(\Omega,\mathcal O)\to(\Omega_\#,\mathcal O_\#)$.
The morphism $\eta$ depends on a cut system $\Gamma$ endowed with
orientations of all cuts $\gamma\in\Gamma$.  $\Omega_\#$ is defined as
contracted cut surface $\Omega/\Gamma$. Stratified surface $\Omega_\#$
inherits special points of $\Omega$ and local orientations at any of them.
The orientations of cuts induce the local orientations at
other special points.

Define {\it  a 'tensor product'}
$\theta: (\Omega',\mathcal O')\times(\Omega'',\mathcal O'')\to
(\Omega,\mathcal O)$ of two pairs $(\Omega',\mathcal O')$ and
$(\Omega'',\mathcal O'')$ as their disjoint union
$(\Omega,\mathcal O)=(\Omega'\sqcup \Omega'',\mathcal O'\sqcup \mathcal O'')$.

Subcategories of the basis category $\mathcal C$ are called {\it
$2D$ categories}. The basic category $\mathcal C$ has subcategories
$\mathcal C_{g,s,\E}$, where $g$ is either an integer or
half-integer nonnegative number or $\infty$, $s$ is an integer
nonnegative number or $\infty$ and $\E=0,1$. The objects of
$\mathcal C_{g,0,0}$ are all pairs $(\OO, \mathcal O)$, where $\OO$
is a stratified surface of type $(\widetilde g, \E, m)$ and
$\widetilde g\leqslant g$. For $s>0$ objects of $\mathcal C_{g,s,0}$
are all pairs $(\OO, \mathcal O)$, where $\OO$ is a stratified
surface of type $(\widetilde g, \E, m, m_1,...,m_{\widetilde s})$,
$\widetilde g \leqslant g$, and $\widetilde s\leqslant
2(g-\widetilde g)+s$. Thus, $\mathcal C=\mathcal
C_{\infty,\infty,0}$. The category $\mathcal C_{g,s,1}$ is  a
subcategory of $\mathcal C_{g,s,0}$. Its objects are all objects
$(\OO, \mathcal O)$ of $\mathcal C_{g,s,0}$ such that $\OO$ is an
orientable surface. Denote by $\mathcal C_{g,s,1,0}$ subcategory of
$\mathcal C_{g,s,1}$, consisting of $(\Omega,\mathcal O)$, where
$\mathcal O$ is generated by some global orientation of $\Omega$.
This global orientation is marked by the same symbol $\mathcal O$.

\subsection{}
Below we define a {\it structure functor}
$(\Omega,\mathcal O) \to V(\Omega,\mathcal O)$
from the basic category of surfaces to the category of
vector spaces \cite{Al-Nat}.

Let $\{X_m | m\in M\}$ be a finite set of $n=|M|$  vector spaces
$X_m$. The action of the symmetric group $S_n$ on the set
$\{1,\dots,n\}$ induces its action on  the linear space
$\left(\oplus_{\sigma:\{1,\dots,n\}\leftrightarrow M}
X_{\sigma(1)}\otimes\dots\otimes X_{\sigma(n)}\right)$, an element
$s\in S_n$ brings a summand $X_{\sigma(1)}\otimes\dots\otimes
X_{\sigma(n)}$ to the summand $X_{\sigma(s(1))}\otimes\dots \otimes
X_{\sigma(s(n))}$. Denote by $\otimes_{m\in M} X_m$ the subspace of
all invariants of this action.

The vector space  $\otimes_{m\in M} X_m$ is canonically isomorphic to a
tensor product of all $X_m$ in any fixed order, the isomorphism
is a projection of the vector space $\otimes_{m\in M} X_m$ to the summand
that is equal to the tensor product of $X_m$ in a given order.
Assume that all $X_m$ are equal to a fixed vector space $X$.
Then any bijection $M\leftrightarrow M'$ of sets induces the isomorphism
$\otimes_{m\in M} X_m\leftrightarrow \otimes_{m'\in M'} X_{m'}$.

Let $A$ and $B$  be finite dimensional vector spaces over a field $\mathbb K$
endowed with involutive linear transformations $A\to A$ and $B\to B$,
which we denote by $x\mapsto x^*$ ($x\in A$) and $y\mapsto y^*$ ($y\in B$) resp.

Let $(\Omega,\mathcal O)$ be a pair, consisting of a stratified surface $\Omega$
and a set $\mathcal O$ of local orientations at its special points.
Denote by $\Omega_a$ the set of all interior special points
and by $\Omega_b$ the set of all boundary special points.
Put also $\Omega_0=\Omega_a\sqcup\Omega_b$.
Assign a copy $A_p$ of a vector space $A$ to any point $p\in\Omega_a$ and
a copy $B_q$ of a vector space $B$ to any point $q\in\Omega_b$.
Put $V(\OO, \mathcal O)=V_{\Omega}=(\otimes_{p\in \Omega_a} A_p)\otimes
(\otimes_{q\in\Omega_b} B_q)$.

The group $\Sigma(\Omega)$ of transpositions of $\Omega_b$ acts naturel on
$V_{\Omega}$.

For any morphism of pairs $(\Omega,\mathcal O)\to (\Omega',\mathcal
O')$ define a morphism of vector spaces $V_{\Omega}\to V_{\Omega'}$
as follows:

1) An isomorphism $\phi:(\Omega,\mathcal O)\to(\Omega',\mathcal O')$
induces the isomorphism $\phi_*:V_{\Omega}\to V_{\Omega'}$ because
$\phi$ generates the bijections $\Omega_a\leftrightarrow\Omega_a'$
and $\Omega_b\leftrightarrow\Omega_b'$ of sets of special points.

2) For a change of local orientations
$\psi:(\Omega,\mathcal O)\to(\Omega,\mathcal O')$
define a linear map $\psi_*:V_{\Omega}\to V_{\Omega}$
as $(\otimes_{r\in \Omega_0} \psi_r)$,
where for any $r\in \Omega_0$

$$\psi_r(x)=
    \left\{\begin{array}{l}
              x, \mbox{\ \ if $o_r=o_r'$}\\
        x^*, \mbox{\ \ if $o_r=-o_r'$}\\
     \end{array}
  \right.\ \ $$

3) In order to define
a morphism $\eta_*:V_{\Omega}\to V_{\Omega_\#}$
for any cutting morphism $\eta:(\Omega,\mathcal O)\to(\Omega_\#,\mathcal O_\#)$
we need to fix elements $\widehat{K}_{A,*}\in A\otimes A$,
$\widehat{K}_{B,*}\in B\otimes B$ and $U\in A$.
(The notation will be clear from the sequel.)

Evidently, it is sufficient to define $\eta_*$ for an arbitrary
oriented cut $\gamma\subset \Omega$. In this case  we have
a canonical isomorphism
$V_{\Omega_\#}=V_{\Omega}\otimes X$, where

$$X=
    \left\{ \begin{array}{l}
      A\otimes A, \mbox{\ \ if $\gamma$ is a coorientable contour,}\\
      B\otimes B, \mbox{\ \ if  $\gamma$ is a segment,}\\
       A, \mbox{\ \ if $\gamma$ is a noncoorientable cut.}
  \end{array}
        \right.\ \ $$

For $x\in V_{\Omega}$ put $\eta_*(x)=x\otimes z$, where $z$ is
either $\widehat{K}_{A,*}$, or $\widehat{K}_{B,*}$, or $U$ resp.

Finally, for a 'tensor product' $\theta:(\Omega',\mathcal O')\times
(\Omega'',\mathcal O'')\to (\Omega'\sqcup\Omega'',\mathcal O'\sqcup
\mathcal O'')$ there is evident canonical linear map $\theta_*:
V_{\Omega'}\otimes V_{\Omega''}\to V_{\Omega'\sqcup\Omega''}$.

\section{ Topological Field Theory over a functor}

\subsection{}
Let $\mathcal R$ be a category of triples $(W,\rho,\Sigma)$ where
$W$ is a vector spaces over a field $\mathbb K$, $\Sigma$ is a
group, and $\rho:\Sigma\to Aut(W)$ is a homomorphism. Morphisme
$(W,\rho,\Sigma)\to (W',\rho',\Sigma')$ is a pair of isomorphisms
$(\vartheta_W: W\to W',\vartheta_{\Sigma}: \Sigma\to \Sigma')$ such
that $\vartheta_W\rho=\rho'\vartheta_{\Sigma}\vartheta_W$.

Let $\mathcal T$ be a functor from a $2D$ category to $\mathcal R$
such that $\mathcal T(\OO, \mathcal O)=
(W(\OO, \mathcal O),\rho(\OO, \mathcal O),\Sigma(\OO, \mathcal O))$,
where $\Sigma(\OO, \mathcal O)=\Sigma(\OO)$ is the group of transpositions
of $\OO_b$

We consider that $\mathcal T(\OO, \mathcal O)=
(\mathbb K,\rho,\Sigma)$, where $\rho(\Sigma)$ is the identical map,
if $\OO$ is a trivial stratified surface, and $\mathcal T(\pi)$
is the identical morphism, if $\pi$ is a morphism of trivial
stratified surfaces.

{\it A Topological Field Theory over $\mathcal T$} is a set
$$\mathcal F=\{A, x\mapsto x^*, B, y\mapsto y^*, \mathcal T,
\Phi_{(\OO, \mathcal O)}\},$$ where $(A, x\mapsto x^*)$ and
$(B, y\mapsto y^*)$ are finite dimensional vector spaces over $\mathbb K$
endowed with involute linear transformation and
$\{\Phi_{(\OO, \mathcal O)}\}$ is a family of linear operators
$\Phi_{(\OO, \mathcal O)}:V_\OO\to W(\OO, \mathcal O)$, where
$V_\OO$ is the image of $(\OO,\mathcal O)$ by the structural
functor for $\{A, x\mapsto x^*, B, y\mapsto y^*\}$
and $\mathcal T(\OO, \mathcal O)=(W(\OO, \mathcal O),\rho(\OO, \mathcal O),
\Sigma(\OO, \mathcal O))$.

The set $\mathcal F$ is called a Topological Field Theory, if
the following axioms are satisfied

$0^\circ$ {\it Algebraic invariance}

For any $\sigma\in\Sigma(\OO, \mathcal O)$ it is required that
$$\Phi_{(\Omega,\mathcal O)}(\sigma(x))=
\rho(\OO, \mathcal O)(\sigma)(\Phi_{(\Omega,\mathcal O)}(x)).$$

$1^\circ$ {\it Topological invariance}

For any isomorphism
of pairs  $\phi:(\Omega,\mathcal O) \to (\Omega', \mathcal O')$ it is
required that
$$\Phi_{(\Omega',\mathcal O')}(\phi_*(x))=
\mathcal T(\phi)\Phi_{(\Omega,\mathcal O)}(x).$$

$2^\circ$ {\it Invariance of a change of  local orientations.}

For any change
of local orientations $\psi:(\Omega,\mathcal O)\to(\Omega,\mathcal O')$
it is required that
$$\Phi_{(\Omega,\mathcal O')}(\psi_*(x))=\mathcal T(\psi)\Phi_{(\Omega,\mathcal O)}(x).$$

$3^\circ$ {\it  Nondegeneracy.}

Define first a bilinear form $(x,x')_A$ on the vector space $A$. Namely,
let $(\Omega,\mathcal O)$ be a pair, where $\Omega$
is a sphere  with 2 interior special points $p,p'$ and the set
    $\mathcal O=\{o_p,o_{p'}\}$ is such that
local orientations $o_p$, $o_{p'}$ induce the same global orientations of
the sphere. Put $(x,x')_A=\Phi_{(\Omega,\mathcal O)}(x_p\otimes x'_{p'})$,
where $x_p$ and $x'_{p'}$ are images of $x\in A$ and $x'\in A$ in $A_p$
and $A_{p'}$ resp. The correctness of this definition follows from axioms
$1^\circ$ and $2^\circ$. Evidently, $(x,x')_A$  is a symmetric bilinear form.

Similarly, define a bilinear form $(y,y')_B$ on the vector space $B$, using
a disc  with 2 boundary special points $q,q'$ instead of a sphere with two
interior special points $p,p'$. As in the previous case, local orientations
$o_q$, $o_{q'}$ must induce the same global orientations of the disc.
Evidently, $(y,y')_B$  is a symmetric bilinear form.

It is required that  forms $(x,x')_A$ and $(y,y')_B$ are nondegenerate.

$4^\circ$ {\it Cut invariance.}

Axioms $1^\circ-3^\circ$ allows us to choose elements
$\widehat{K}_{A,*}\in A\times A$, $\widehat{K}_{B,*}\in B\times B$
and $U\in A$. Indeed, any nondegenerate bilinear form on the vector
space $X$ canonically defines [\cite{Al-Nat}, Section 2] the tensor
Casimir element $\widehat{K}_X\in X\otimes X$. Taking forms
$(x,x')_{A,{*}}=(x,{x'}^{*})_A$ and $(y,y')_{B,*}=(y,{y'}^{*})_B$,
we obtain elements $\widehat{K}_{A,*}$ and $\widehat{K}_{B,*}$.

A linear form $\Phi_{(\Omega,\mathcal O)}$ for a projective plane $\Omega$
with one interior special point is an element of the vector space
dual to $A$. We denote by $U$ the image of this element
in $A$ under the isomorphism induced by nondegenerate bilinear form
$(x,x')_A$.

We shall use just these elements for morphisms $\eta_*$ of type 3).

For any cut system $\Gamma$ endowed with orientations of all cuts
it is required that
$$\Phi_{(\Omega_\#,\mathcal O_\#)}(\eta_{*}(x))=
\mathcal T(\eta)\Phi_{(\Omega,\mathcal O)}(x).$$

$5^\circ$  {\it Multiplicativity.}

For the product $\theta:(\Omega',\mathcal O')\times
(\Omega'',\mathcal O'')\to
(\Omega'\sqcup\Omega'',\mathcal O'\sqcup \mathcal O'')$ of
 any two pairs $(\Omega',\mathcal O')$ and $(\Omega'',\mathcal O'')$
it is required that
$$\Phi_{(\Omega_1\sqcup\Omega_2,\mathcal O_1 \sqcup \mathcal O_2)}
(\theta_*(x_1\otimes x_2))=
\Phi_{(\Omega_1,\mathcal O_1)}(x_1)\otimes
\Phi_{(\Omega_2,\mathcal O_2)}(x_2).$$

\subsection{}
Let us consider now some example of Topological Field
Theory on $2D$ categories.
In this subsection we consider only functors
$\mathcal T(\OO, \mathcal O)=
(W(\OO, \mathcal O),\rho(\OO, \mathcal O),\Sigma(\OO, \mathcal O))$,
such that $\rho(\OO, \mathcal O)(\sigma)$ is the identical map for all
$\sigma\in\Sigma(\OO, \mathcal O)$.

1) {\it Topological Field Theory on $\mathcal C_{g,s,\E}$}
(over trivial functor).

Consider the functor, corresponding the field $\mathbb K$ to all objects
$(\OO, \mathcal O)$ and corresponding identical map to any morphism.

a) This gives {\it an Atiayh--Witten $2D$ Topological Field
Theory} \cite{At}, \cite{Wit} for the category
$\mathcal C_{\infty,0,1}$ and involution $x^*=x$.

b) The category $\mathcal C_{\infty,\infty,1}$ and the involutions
$x^*=x, y^*=y$ gives '{\it an Open--Closed' Topological Field
Theory} in the sense of Lazaroiu \cite{Laz}.

c) The category $\mathcal C_{\infty,\infty,\E}$ gives
{\it a Klein Topological Field Theory} in the sense of \cite{Al-Nat}
(without units) .

2) {\it Cohomological Field Theories on $\mathcal C_{g,s,\E}$.}

Recall that a Klein surface of type $(g,s, \E)$ is a surface
of type $(g,s,\E)$, endowed with a dianalytic structure, i.e.,
an atlas with holomorphic and antiholomorphic transitions
functions \cite{Al-Gr}. It is equivalent to a real algebraic
curve (for an information about real algebraic
curves see \cite{Nat2}). A Klein surface of type
$G=(g, \E, m, m_1,...,m_s)$ is called a Klein surface with
special points of type $G$. The moduli space of Klein surfaces is
constructed in \cite{Nat1}.
Let $H^*(\bar M_{G}, \mathbb K)$ be a cohomological algebra of
Deline--Mumford compactification of the space of Klein surfaces
of type $G$. An identification of special points gives some embeddings
$\overline M_{G_1}\times \overline M_{G_2}\to \overline M_{G}$ and
$\overline M_{G_1}\to \overline M_{G}$ by analogy with \cite{Man}.

Cohomological Field Theory in our conception  is defined
as the Topological Field Theory over
the functor $\mathcal T$ that associate
the algebra $H^*(\bar M_{G}, \mathbb K)$ with each object
$(\OO, \mathcal O)$, where $G$ is the type of $\OO$, and associate
the homomorphisms generated by the embeddings
$\overline M_{G_1}\times \overline M_{G_2}\to \overline M_{G}$
and $\overline M_{G_1}\to \overline M_{G}$ with the cutting
morphisms. Here $x^*=x$, $y^*=y$ and the other morphisms are
the same as that for the trivial functor.

This definition takes the (complete) Cohomological Field Theory
as regards to \cite{Man} for subcategories $\mathcal C_{0,0,1}$
($\mathcal C_{\infty,0,1}$).

\subsection{}
According to \cite{Man}, the Cohomological Field Theory over
$\mathcal C_{0,0,1}$ generate some deformations of Atiayh--Witten
$2D$ Topological Field Theory in genus 0. These deformations are
described by formal Frobenius manifolds, i.e., formal solutions of
WDVV equations \cite{Dij, Wit2}.

Our goal is the construction of a functor on $\mathcal C^0_{0,1,1}$
such that Topological Field Theory over this functor generate
deformations of Open--Closed Topological Field Theories in genus 0.
We shall prove that these deformations are described by formal
solutions of  some noncommutative analogues of WDVV equations.

\section {Stable Field Theories}

\subsection{}
Let $\Omega=(\Omega, \mathcal O)$ be a oriented stratified surface
and $\mathcal O$ be generated by some orientation. (Here and later
we omit the mark of orientation $\mathcal O$, if it is constant in
the construction). The orientation of $\OO$ gives a natural sense to
inequalities $b_1<b_2<b_3$ for points from a connected component of
the boundary of $\Omega$. For $C\subset\Omega_b$ denote by
$\Sigma^\circ(C)$ the group of transpositions $\sigma\in
\Sigma(\Omega)$, such that  $\sigma(b_1)< \sigma(b_2)<\sigma(b_3)$
if $b_1< b_2< b_3$ and $b_1, b_2, b_3\in\partial\OO$.

A pair $(\Gamma,\sigma)$, where $\Gamma\subset\Omega$ is a cut
system and $\sigma\in \Sigma(\Omega)$ is called {\it a tiling}, if
all connected components of $\Omega/\Gamma$ are nontrivial. A tiling
$(\Gamma,\sigma)$, where $\Gamma=\emptyset$, is also assumed and it
is called {\it empty tiling}.

Two tiling $(\Gamma',\sigma')$ and $(\Gamma'',\sigma'')$ are called
{\it isomorphic}, if there exists a homeomorphism
$\F:\Omega\to\Omega$, preserving the orientation, moving $\sigma'$
to $\sigma''$ and such that $\F(\GG')=\GG''$. An isomorphic class of
a tiling is called {\it a diagram}. Let $[\Gamma,\sigma]$ be the
diagram corresponding to a tiling $(\Gamma,\sigma)$.

Denote by $\mathbb K(\OO)$ the vector space over $\mathbb K$ generated by all
diagrams of $\Omega$.

Consider a tiling $(\Gamma,\sigma)$, corresponding to a diagram
$T\in \mathbb K(\OO)$. A connected component of $\OO\setminus
\Gamma$ is called {\it a vertex} of the tiling. Denote by $V(T)$ the
set of vertices of $T$.

Let $\OO^T_v$ be the connected component of $\OO/\Gamma$,
corresponding to a vertex $v$. Let $A^T_v$ and $B^T_v$ be the set of
interior and boundary special points of $\OO^T_v$. We can associate
$\sigma^*\in \Sigma(\Omega)$ to any
$\sigma^T_v\in\Sigma(\Omega^T_v)$, considering that
$\sigma^*(b_1)<\sigma^*(b_2)<\sigma^*(b_3)$ if
$b_1,b_2,b_3\notin\OO^T_v $ belong to the same connected component
of $\OO\setminus \Gamma$ and $b_1< b_2< b_3$. Moreover, we  can
consider any cut system on $\Gamma^T_v\in\OO^T_v$ as a cut system on
$\Omega$. Thus, a diagram $S=(\Gamma^T_v,\sigma^T_v)\in \mathbb
K(\OO^T_v)$ generates the diagram $(T,v,S)=(\Gamma\cup\Gamma^T_v,
\sigma^*\sigma)\in(\mathbb K(\OO))$.

\subsection{}
A tiling $(\Gamma,\sigma)$ is called {\it simple}, if $\Gamma$
consist of one cut. A simple tiling is called {\it $A$-tiling}, if
$\Gamma$ is a closed contour homotopic to zero, {\it $B$-tiling}, if
$\Gamma$ is a segment, dividing the surface, {\it $C$-tiling}, if
$\Gamma$ is a closed contour nonhomotopic to zero, and {\it
$D$-tiling}, if $\Gamma$ is a segment, nondividing the surface.
Diagrams, corresponding to $A (B, C, D)$, {\it empty} tilings are
called {\it $A (B, C, D)$  empty diagram}.

Let $C_1$ and $C_2$ be subsets of a surface $\OO$. By $r_A(C_1|C_2)$
(respectively, $r_B(C_1|C_2)$) denote the set of $A-$ (respectively,
$B-$) diagrams, corresponding to $(\Gamma,\sigma)$, where $\Gamma$
divide $C_1$ and $C_2$.

For a sphere $\Omega$ and $a_1,a_2,a_3,a_4\in\Omega_a$ put
$r_{\Omega}(a_1,a_2|a_3,a_4)= r_A(a_1, a_2|a_3, a_4)$.

Let $\Omega$ be a disk, $b_1,b_2,b_3,b_4\in\Omega_b$ and
$b_1<b_2<b_3<b_4$. Denote by $r_{\Omega}(b_1,b_2|b_3,b_4)$ the set
of all diagrams, belonging to $r_B(\sigma(b_1)\cup \sigma(b_2)|
\sigma(b_3), \cup \sigma(b_4))$ and corresponding to $(\Gamma,
\sigma)$, where $\sigma\in \Sigma^\circ(\sigma(b_1)\cup
\sigma(b_2)\cup \sigma(b_3)\cup \sigma(b_4))\cap\Sigma(\OO_b -
(\sigma(b_1)\cup \sigma(b_2)\cup \sigma(b_3)\cup \sigma(b_4)))$.

Let $\Omega$ be a disk  $a\in\Omega_a$, $b_1,b_2\in\Omega_b$,
$\sigma\in \Sigma(\Omega)$. Denote by $r_{\Omega}(b_1,b_2|a)$ the
set of all diagrams belonging to $r_B(\sigma(b_1)\cup
\sigma(b_2)|a)$ and corresponding to $(\Gamma, \sigma)$, where
$\sigma\in \Sigma(\Omega_b-(\sigma(b_1)\cup\sigma(b_2)))$ and any
ends $b$ of $\Gamma$ satisfy $\sigma(b_1)<b<\sigma(b_2)$.

Let $\Omega$ be a disk  $a_1,a_2\in\Omega_a$, $b\in\Omega_b$. Put
$r_{\Omega}(a_1,a_2|b)=r_A(a_1, a_2|b)$. Denote by
$r_{\Omega}(a_1,b|a_2)r_{\Omega}(a_2,b|a_1)$ the diagrams
corresponding to $(\Gamma_1\cup \Gamma_2,\sigma)$, where
$(\Gamma_1,\sigma)$ from $r_B(a_1\cup\sigma(b)|a_2), (\Gamma_2,
\sigma'')$ from $r_B(a_2\cup\sigma(b)|a_1)$, and $\sigma\in \Sigma
(\Omega_b-b)$.

Let $\Omega$ be a cylinder. Denote by $r^\sigma_\Omega(C)$ and
$r^\sigma_\Omega(D)$ the set of all $C-$ and $D-$diagrams
respectively.

Denote by $H^*(\Omega)$ the set of linear functions $l:\mathbb
K(\OO)\to \mathbb K$, that is equal 0 on all elements in the forms:

1) $$[\phi(\Gamma),\sigma(\phi)\sigma]-[\Gamma,\sigma]$$ for any
tiling $(\Gamma,\sigma)$ of $\Omega$, any isomorphism
$\phi:(\Omega,\mathcal O)\to(\Omega,\mathcal O)$ and the
transposition $\sigma(\phi)=\phi|_{\OO_b}$

2) $$\sum\limits_{\begin{array}{c} S\in
r_{\Omega^T_v}(a_1,a_2|a_3,a_4)\end{array}} (T,v,S) -
    \sum\limits_{\begin{array}{c}
S\in r_{\Omega^T_v}(a_2,a_3|a_4,a_1)\end{array}} (T,v,S)$$ for any
diagram $T\in\mathbb K(\OO)$, any $v\in V(T)$ such that
$\Omega^T_v$, is a sphere, and any pairwise different points
$a_1,a_2,a_3,a_4\in(\Omega^T_v)_a$.

3) $$\sum\limits_{\begin{array}{c} S\in
r_{\Omega^T_v}(b_1,b_2|b_3,b_4)\end{array}} (T,v,S) -
    \sum\limits_{\begin{array}{c}
S\in r_{\Omega^T_v}(b_2,b_3|b_4,b_1)\end{array}} (T,v,S)$$ for any
diagram $T\in\mathbb K(\OO)$, any $v\in V(T)$ such that
$\Omega^T_v$, is a disk, and any $b_1,b_2,b_3,b_4\in(\Omega^T_v)_b$
such that $b_1<b_2 <b_3<b_4$.

4) $$\sum\limits_{\begin{array}{c} S\in
r_{\Omega^T_v}(b_1,b_2|a)\end{array}} (T,v,S) -
    \sum\limits_{\begin{array}{c}
S\in r_{\Omega^T_v}(b_2,b_1|a)\end{array}} (T,v,S)$$ for any diagram
$T\in\mathbb K(\OO)$, any $v\in V(T)$ such that $\Omega^T_v$,is a
disk, any $a\in(\Omega^T_v)_a$, and any  different points
$b_1,b_2\in(\Omega^T_v)_b$.

5) $$\sum\limits_{\begin{array}{c} S\in
r_{\Omega^T_v}(a_1,a_2|b)\end{array}} (T,v,S) -
    \sum\limits_{\begin{array}{c}
S\in r_{\Omega^T_v}(a_1,b|a_2)
    r_{\Omega^T_v}(a_2,b|a_1)\end{array}} (T,v,S)$$
for any diagram $T\in\mathbb K(\OO)$, any $v\in V(T)$ such that
$\Omega^T_v$, is a disk with $n+2$ interior and $m+1$ boundary
marked points,  $a\in (\Omega^T_v)_a$, and distinct
$b_1,b_2\in(\Omega^T_v)_b$.

6) $$\sum\limits_{\begin{array}{c} S\in
r_{\Omega^T_v}(C)\end{array}} (T,v,S) -
    \sum\limits_{\begin{array}{c}
S\in r_{\Omega^T_v}(D)\end{array}} (T,v,S)$$ for any diagram
$T\in\mathbb K(\OO)$, and any $v\in V(T)$ such that $\Omega^T_v$, is
a cylinder with $n$ interior marked points, with $m+2$ boundary
marked points and also any boundary contour contained marked points.

Define the action $\rho(\Omega):\Sigma(\Omega)\to Aut(H^*(\Omega))$ by
$\sigma(l)(\Gamma,\sigma')=l(\Gamma,\sigma'\sigma)$.

Extend our definitions  of $\mathbb K(\OO)$, $H^*(\OO)$ and $\rho(\OO)$
on disconnected sums $\OO=\OO_1\coprod\OO_2$ by $\mathbb K(\OO)=
\mathbb K(\OO_1)\otimes\mathbb K(\OO_2)$, $H^*(\OO)=H(\OO_1)^*
\otimes H(\OO_2)^*$. The actions $\rho(\Omega_1)$ and $\rho(\Omega_2)$
generate the action $\rho(\Omega_1)$. This gives a possible to define
$H^*(\OO)$ and $\rho(\Omega)$ for any disconnected sums of oriented sphere
and disks.

Let $\gamma$ be a simple cut on $\Omega$ and $\OO_\#=\OO/\G$.
Denote by $\mathbb K_{\gamma}(\OO)\subset\mathbb K(\OO)$  the subspace
generated by all diagrams containing the cut $\gamma$.
A diagram $T\in \mathbb K(\OO_\#)$ generates a diagram
$\widetilde T\in\mathbb K_{\gamma}(\OO)$. This correspondence defines a
monomorphism $\varepsilon_\gamma:\mathbb K(\OO_\#)\to \mathbb K_\gamma(\OO)
\subset K(\OO)$ and thus a homomorphism $\varepsilon^*_\gamma:H^*(\OO)
\to H^*(\OO_\#)$.

\subsection {}
Now we introduce {\it a stabilising functor} $\mathcal F$ on the
category $\mathcal C^0_{0, 1, 1}$ with structure functor, generated
by arbitrary data $\{A, x\mapsto x^*, B, y\mapsto y^*\}$.

For local orientations $\mathcal O$, generated by a global
orientation we put $\mathcal F(\OO,\mathcal O)=
(H^*(\OO),\rho(\OO),\Sigma(\OO))$.

For isomorphism $\phi:(\OO,\mathcal O)\to (\OO', \mathcal O')$
denote by  $\mathcal F(\phi):H^*(\Omega,\mathcal O)
\to H^*(\Omega', \mathcal O')$, the isomorphism that it generates.

Chang of global orintation $\psi:(\Omega,\mathcal O)
\to(\Omega,\mathcal O')$ is generated by an isomorphism
$\phi:(\OO,\mathcal O)\to (\OO', \mathcal O')$. Put $\mathcal F(\psi)=
\mathcal F(\phi)$.

For cutting morphism $\eta:(\OO,\mathcal O)\to (\OO_\#,\mathcal
O_\#)$ by a cut $\gamma\subset \OO$ we consider that $\mathcal
F(\eta)$ is generated by $\varepsilon^*_\gamma$.

By definition, Topological Field Theory $$\mathcal N=\{A, x\mapsto
x^*, B, y\mapsto y^*, \mathcal F, \Phi_{(\Omega,\mathcal O)}\}$$
with values in a stabilizing
functor $\mathcal F$ is called a {\it Stable Field Theory}.

{\bf Remark.} It follows from \cite{Man} that the codiagram space
$H^*(\Omega)$ of a sphere with $n$ special points coincides with the
cohomological algebra $H^*(\bar M_{0,n}, \mathbb K)$ of the
Deligne-Mumford compactification of moduli space of spheres with $n$
punctures. Thus, for $x^*=x$ Stable Field Theory on $2D$ category
$\mathcal C_{0,0,1}$ coincides with the Kontsevich-Manin
Cohomological Field Theory \cite{Kon-Man}.

\section {System of disk correlation functions}

\subsection{}
Let $A$ and $B$ be vector spaces with basises
$\{\alpha_1,...,\alpha_n\}\subset A,\ \{\beta_1,...,\beta_m\}\subset B$
and involutions $x\mapsto x^*, y\mapsto y^*$ ($x, x^*\in A,
y,y^*\in B$). Consider a collection of tensors
$\{f_{r,\ell}:A^{\otimes r}\otimes B^{\otimes\ell}\to \mathbb K\}$. By
$<x_1,...,x_k, y_1,...,y_\ell>$ denote $f_{k,\ell}(x_1\otimes\dotsb
\otimes x_k\otimes y_1\otimes\dotsb \otimes y_\ell)$, where
$x_i\in A, y_i\in B$. We say that this collection is {\it a System of
Disk Correlation Function}, if the following conditions hold.

{\bf Axiom $1^0$:} The value $<x_1,...,x_k, y_1,...,y_\ell>$
is invariant under any permutation of $x_i$ and cyclic permutation of $y_i$.

{\bf Axiom $2^0$:}  $<x_1,...,x_k, y_1,...,y_\ell>=
<x_1^*,...,x_k^*, y_\ell^*,y_{\ell-1}^*,...,y_1^*>$.

{\bf Axiom $3^0$:} The bilinear forms $<x_i,x_j>$ and
$<y_i,y_j>$ are nondegenerate. Denote by $F^{\alpha_i\cdot\alpha_j}$
and $F^{\beta_i\cdot\beta_j}$ the inverse matrices for
$F_{\alpha_i\alpha_j}=<\alpha_i,\alpha_j>$ and $
F_{\beta_i\beta_j}=<\beta_i,\beta_j>$ .

{\bf Axiom $4^0$:}  Let  $x_1, x_2, x_3, x_4,x^i \in A$ and
$\mathcal A=\{x^1,...,x^r\}$. Denote by $<\mathcal
A|x_1,x_2|x_3,x_4>$ the sum of all numbers as
$<x_1,x_2,x^{\SI(1)},...,x^{\SI(s)}, \alpha_i>
F^{\alpha_i\alpha_j}<\alpha_j,x_3, x_4,
x^{\SI(s+1)},...,x^{\SI(r)}>$, where $0\leqslant s\leqslant r$,
$1\leqslant i, j\leqslant n$ and $\SI$ passes through all
permutations of $r$ indices.

The axiom says that $$<\mathcal A|x_1, x_2|x_3,x_4>=
<\mathcal A|x_4,x_1|x_2,x_3>$$
for any $\mathcal A$ and $x_1,x_2,x_3,x_4$.

{\bf Axiom $5^0$:}  Let $x^i\in A$, $\mathcal A=\{x^1,...,x^r\}$,
$y_1,y_2,y_3,y_4,y^i\in  B$, and $\mathcal B=\{y^1,...,y^p\}$ .
Denote by $<\mathcal A,\mathcal B|y_1,y_2|y_3,y_4>$ the sum of all
numbers as $<x^{\SI(1)},...,x^{\SI(s)},
y^{\xi(1)},...,y^{\xi(p_1)},y_1,y^{\xi(p_1+1)},...,y^{\xi(p_2)},y_2,
y^{\xi(p_2+1)},...,y^{\xi(p_3)},\beta_i>F^{\beta_i\beta_j}
<\beta_j,x^{\sigma(s+1)},...,x^{\sigma(r)},y^{\xi(p_3+1)},...,y^{\xi(p_4)},y_3,
y^{\xi(p_4+1)},...,y^{\xi(p_5)},y_4$,
$y^{\xi(p_5+1)},...,y^{\xi(p)}>$, where $0\leqslant s\leqslant r$,
$0\leqslant p_1\leqslant p_2\leqslant p_3 \leqslant p_4\leqslant
p_5\leqslant p$, $1\leqslant i,\ j\leqslant m$, $\SI$ passes through
all permutations of $r$ indices and $\xi$ passes through all
permutations of $p$ indices.

The axiom says that $$<\mathcal A,\mathcal B|y_1,y_2|y_3,y_4>=
<\mathcal A,\mathcal B|y_4,y_1|y_2,y_3>$$
for any $\mathcal A$,$\mathcal B$, and $y_1,y_2,y_3,y_4$.

{\bf Axiom $6^0$:}  Let $x,x^i\in A$, $\mathcal A=\{x^1,...,x^r\}$,
$y_1,y_2,y^i\in  B$, and $\mathcal B=\{y^1,...,y^p\}$ Denote by
$<\mathcal A,\mathcal B|x|y_1,y_2>$ the sum of all numbers as
$<x,x^{\SI(1)},...,x^{\SI(s)},
y^{\xi(1)},...,y^{\xi(p_1)},\beta_i>F^{\beta_i\beta_j}
<\beta_j,x^{\sigma(s+1)},...,x^{\sigma(r)},y^{\xi(p_1+1)},...,y^{\xi(p_2)},y_1,
y^{\xi(p_2+1)},...,y^{\xi(p_3)},y_2,y^{\xi(p_3+1)},...,$
$y^{\xi(p)}>$, where $0\leqslant s\leqslant r$, $0\leqslant
p_1\leqslant p_2\leqslant p_3\leqslant p$, $1\leqslant i,\
j\leqslant m$, $\SI$ passes through all permutations of $r$ indices
and $\xi$ passes through all  permutations of $p$ indices.

The axiom says that $$<\mathcal A,\mathcal B|x|y_1,y_2>= <\mathcal
A,\mathcal B|x|y_2,y_1>$$ for any $\mathcal A$,$\mathcal B$, $x$,
and $y_1, y_2$.

{\bf Axiom $7^0$:}  Let $x_1,x_2,x^i\in A$, $\mathcal
A=\{x^1,...,x^r\}$, $y,y^i\in  B$, and $\mathcal B=\{y^1,...,y^p\}$.
Denote by $<\mathcal A,\mathcal B|x_1,x_2|y>$ the sum of all numbers
as $<x_1,x_2,x^{\SI(1)},...,x^{\SI(s)},\alpha_i>
F^{\alpha_i\alpha_j}<\alpha_j,x^{\sigma(s+1)},...,x^{\sigma(r)},
y,y^{\xi(1)},...,y^{\xi(p)}>$, where $0\leqslant s\leqslant r$
$1\leqslant i,\ j\leqslant n$, $\SI$ passes through all permutations
of $r$ indices and $\xi$ passes through all  permutations of $p$
indices.

Denote by $<\mathcal A,\mathcal B|x_1,y|x_2>$ $<\mathcal A,\mathcal
B|x_2,y|x_1>$ the sum of all numbers as
$<x,x^{\SI(1)},...,x^{\SI(s)}, y,y^{\xi(1)},...,y^{\xi(p_1)},$
$\beta_i^1,y^{\xi(p_2+1)},...,y^{\xi(p_3)},$ $\beta_i^2,
y^{\xi(p_4+1)}, ..., y^{\xi(p)}>F^{\beta_i^1\beta_j^1}
F^{\beta_i^2\beta_j^2} <\beta_j^1, x^{\sigma(s+1)},
...,x^{\sigma(t)}, y^{\xi(p_1+1)},...,y^{\xi(p_2)}> <\beta_j^2,
x^{\sigma(t+1)},...,x^{\sigma(r)},
y^{\xi(p_3+1)},...,y^{\xi(p_4)}>$, where $0\leqslant s\leqslant r$,
$0\leqslant p_1\leqslant p_2\leqslant p_3\leqslant p_4\leqslant p$,
$1\leqslant i^1,\ j^1,\ i^2,\ j^2\leqslant m$, $\SI$ passes through
all permutations of $r$ indices and $\xi$ passes through all
permutations of $p$ indices.

The axiom says that $$<\mathcal A,\mathcal B|x_1,x_2|y>=
<\mathcal A,\mathcal B|x_1,y|x_2><\mathcal A,\mathcal B|x_2,y|x_1>$$
for any $\mathcal A$,$\mathcal B$, $x_1,x_2$, and $y$.

We say that a system of disk correlation is {\it extended} if the
following condition holds.

{\bf Axiom $8^0$:}  Let $x^i\in A$, $\mathcal A=\{x^1,...,x^r\}$,
$y_1,y^i_2\in B$, and $\mathcal B=\{y^1,...,y^p\}$. Denote by
$<\mathcal A,\mathcal B|y_1,y_2>_a$ the sum of all numbers as
$<x^{\SI(1)},...,x^{\SI(s)}, y_1,y^{\xi(1)},...,y^{\xi(q)},\alpha_i>
F^{\alpha_i\alpha_j}<\alpha_j,x^{\sigma(s+1)},...,x^{\sigma(r)},
y_2,y^{\xi(q+1)},...,y^{\xi(p)}>$, where $0\leqslant s\leqslant r$,
$0\leqslant q\leqslant p$, $1\leqslant i,\ j\leqslant n$, $\SI$
passes through all permutations of $r$ indices and $\xi$ passes
through all permutations of $p$ indices.

Denote by $<\mathcal A,\mathcal B|y_1,y_2>_b$ the sum of all numbers
as $<x^{\SI(1)},...,x^{\SI(r)}, y_1, y^{\xi(1)},...,y^{\xi(q_1)},
\beta_i, y^{\xi(q_1+1)}, ..., y^{\xi(q_2)},$ $y_2,
y^{\xi(q_2+1)},..., y^{\xi(q_3)}, \beta_j,$
$y^{\xi(q_3+1)},...,y^{\xi(p)}>F^{\beta_i\beta_j}$, where
$0\leqslant q_1\leqslant q_2\leqslant q_3\leqslant p$, $1\leqslant
i,\ j\leqslant n$, $\SI$ passes through all permutations of $r$
indices and $\xi$ passes through all permutations of $p$ indices.

The axiom says that $$<\mathcal A,\mathcal B|y_1,y_2|y>_a= <\mathcal
A,\mathcal B|y_1,y_2>_b$$ for any $\mathcal A$,$\mathcal B$, $y_1$
and $y_2$.

It is easy to prove that the axioms are fulfilled  for any
bases $\{\alpha_1,...,\alpha_n\}\subset A,
\{\beta_1,...,\beta_m\}\subset B$,
if they are fulfilled for one pair of such bases.

\subsection{}
Let $\mathcal N=\{A,  x\mapsto x^*, B, y\mapsto y^*, \mathcal F,
\Phi_{(\Omega,\mathcal O)}\}$ be a Stable Field Theory on the
category $\mathcal C^0_{0, 1, 1}$ . Consider the collection of
tensors $<x_1,..., x_k,$ $y_1,...,y_\ell>_{\mathcal N}=
\Phi_{(\Omega, \mathcal O)}(z)(\emptyset,1)$. Here $\Omega$ is a
sphere (if $\ell=0$) or a disk (if $\ell>0$) with $k$ special
interior points $a_1,...,a_k$ and with $\ell$ special boundaries
points $b_1,...,b_\ell$, $\emptyset$ is the empty diagram of $\OO$
and $1$ is identical permutation. We assume that the points
$b_1,...,b_\ell$ are ordered by the orientation of disk $\Omega$ and
$z=x_1\otimes\dotsb \otimes x_k\otimes y_1\otimes \dotsb\otimes
y_\ell$, where $x_i$ corresponds to $a_i$ and $y_i$ corresponds to
$b_i$. We say that the tensors $<x_1\dotsb y_\ell>_{\mathcal N}$ are
{\it generated} by $\mathcal N$.

{\bf Theorem 4.1.} {\sl 1) The correspondence $\mathcal N\mapsto
\{<x_1,...y_\ell>_{\mathcal N}\}$ is one-to-one correspondence
between Stable Field Theories  on $\mathcal C_{0,1,1}$ and Systems
of Disk Correlation Functions. If $\mathcal N$ is Stable Field
Theories on $\mathcal C_{0,2,1}$ then $\{<x_1,...,y_/ell>_\mathcal
N$ is an extended systems of disk correlation functions.}

Proof: 1) Prove that $\{<x_1,...,y_{\ell}>_{\mathcal N}\}$ is a
Extended System of Disk Correlation Functions. Axioms  $1^0, 2^0,
3^0$ for Systems of Disk Correlation Functions follow from axioms
$1^\circ, 2^\circ, 3^\circ$ for Topological Field Theories. The
axioms $4^\circ - 7^\circ$ are fulfilled because $\Phi_{(\Omega,
\mathcal O)}(z)\in H^*(\Omega)$.

2) Now let $\{<x_1,...,y_\ell>\}$ be a System of Disk Correlation
Functions. Consider a sphere (if $\ell=0$) or a disk (if $\ell>0$)
$\Omega$ with $k$ special interior points $a_1,...,a_k$ and with
$\ell$ special boundaries points $b_1,...,b_\ell$. Put
$\Phi_{(\Omega, \mathcal O)}(z)(\emptyset,1)= <x_1,..., x_k,$
$y_1,...,y_\ell>_{\mathcal N}$. We assume that the points
$b_1,...,b_\ell$ are ordered by the orientation of disk $\Omega$ and
$z=x_1\otimes\dotsb \otimes x_k\otimes y_1\otimes \dotsb\otimes
y_\ell$, where $x_i$ corresponds to $a_i$ and $y_i$ corresponds to
$b_i$. $\emptyset$ is the empty diagram of $\OO$ and $1$ is
identical permutation.

Using axioms $0^\circ, 4^\circ, 5^\circ$ for Topological Field Theories we can
uniquely  continue $\Phi_{(\Omega, \mathcal O)}(z)$ up to a linear function
$l:\mathbb K(\Omega, \mathcal O)\to\mathbb K$. It is follows from
the axioms  $4^0, 5^0, 6^0, 7^0$ for Systems of Disk Correlation
Functions that $l\in H^*(\Omega, \mathcal O)$. It is follows from axioms
$1^0, 2^0, 3^0$ for Systems of Disk Correlation Functions  that
$1^\circ, 2^\circ, 3^\circ$ for Topological Field Theories are fulfilled.
$\square$

\section{Structure equations}

\subsection{}
Let, as above, $A, x\mapsto x^*$ and $B, y\mapsto y^*$ be vector
spaces with involutions. These involutions generate involutions $*$
on $A^{\otimes k}\otimes B^{\otimes\ell}$ by the rule
$*(x_1\otimes\dotsb\otimes x_k\otimes y_1\otimes\dotsb\otimes y_\ell)=
(x_1^*\otimes\dotsb\otimes x_k^*\otimes y_\ell^*\otimes\dotsb\otimes y^*_1)$.
Fix bases
$\{\alpha_1,...,\alpha_n\}\subset A$ and $\{\beta_1,...,\beta_m\}\subset B$.
By definition, we consider that a multiplication of
monomials $\alpha_{i_1}\otimes\dotsb\otimes \alpha_{i_k}\otimes
\beta_{j_1}\otimes\dotsb\otimes\beta_{j_\ell}$ and
$\alpha_{\widetilde i_1}\otimes\dotsb\otimes
\alpha_{\widetilde i_{\widetilde k}}
\otimes \beta_{\widetilde j_1}\otimes \dotsb
\otimes\beta_{\widetilde j_{\widetilde\ell}}$
is the monomial
$\alpha_{i_1}\otimes\dotsb\otimes \alpha_{i_k}
\otimes\alpha_{\widetilde i_1}\otimes
\dotsb\otimes \alpha_{\widetilde i_{\widetilde k}}
\otimes \beta_{j_1}\otimes
\dotsb \otimes\beta_{j_{\ell}}\otimes \beta_{\widetilde j_1}
\otimes\dotsb\otimes\beta_{\widetilde j_{\widetilde\ell}}$.

Now consider {\it a tensor series}, i.e., a formal series
of tensor monomials $F=\sum c(i_1\dotsb i_k|j_1\dotsb j_\ell)
\alpha_{i_1}\otimes\dotsb\otimes\alpha_{i_k}\otimes\beta_{j_1}
\otimes\dotsb\otimes\beta_{j_\ell}$, where
$c(i_1,...,i_k| j_1,...,j_\ell)\in \mathbb K$.
The involution $*$ natural acts on
the set of such series. The multiplication of monomial
defines the multiplication of formal series.

Define "partial derivative" of tensor series. It is a linear
continuation of partial derivative of monomials.

Let $\partial(\alpha_{i_1}\otimes\dotsb\otimes
\alpha_{i_k}\otimes\beta_{j_1}\otimes\dotsb\otimes\beta_{j_\ell}) /
\partial\alpha_{i}$ be the sum of monomials as $\alpha_{i_1}
\otimes\dotsb\otimes\alpha_{i_{p-1}}\otimes\alpha_{i_{p+1}}\otimes\dotsb\otimes
\alpha_{i_k}\otimes\beta_{j_1}\otimes\dotsb\otimes\beta_{j_{\ell}}$
such that $i_p=i$.

Similarly, let $\partial(\alpha_{i_1}\otimes\dotsb\otimes
\alpha_{i_k}\otimes\beta_{j_1}\otimes\dotsb\otimes\beta_{j_\ell}) /
\partial\beta_{i}$ be the sum of monomials as $\alpha_{i_1}
\otimes\dotsb\otimes\alpha_{i_{k}}\otimes\beta_{j_1}\otimes\dotsb
\otimes\beta_{j_{p-1}}\otimes\beta_{j_{p+1}}\otimes\dotsb\otimes\beta_{j_\ell}$
such that $j_p=i$.

Put $\partial^2/\partial\alpha_i\partial\alpha_j =
(\partial/\partial\alpha_i) (\partial/\partial\alpha_j),$
$\partial^2/\partial\alpha_i\partial\beta_j =
(\partial/\partial\alpha_i) (\partial/\partial\beta),$
$\partial^2/\partial\beta_i\partial\beta_j =
(\partial/\partial\beta_i) (\partial/\partial\beta_j),$
$\partial^3/\partial\alpha_i\partial\alpha_j \partial\alpha_r=
(\partial/\partial\alpha_i) (\partial/\partial\alpha_j)
(\partial/\partial\alpha_r).$

A definition of $\partial^3(\alpha_{i_1}\otimes\dotsb\otimes
\alpha_{i_k}\otimes\beta_{j_1}\otimes\dotsb\otimes\beta_{j_\ell}) /
\partial\beta_{i}\partial\beta_{j}\partial\beta_{r}$ is more
complicated. It ia the sum of monomials as $\alpha_{i_1}
\otimes\dotsb\otimes\alpha_{i_{k}}\otimes\beta_{s_2}\otimes\dotsb
\otimes\beta_{s_{p-1}}\otimes\beta_{s_{p+1}}\otimes\dotsb\otimes\beta_{s_{q-1}}
\otimes\beta{s_{q+1}}\otimes\beta_{s_\ell}$ such that sequences
$\beta_i,\beta_{s_2},...,\beta_{s_{p-1}},\beta_j,\beta_{s_{p+1}},...,
\beta_{s_{q-1}},\beta_r, \beta{s_{q+1}}, \beta_{s_\ell}$ and
$\beta_{j_1},...,\beta_{j_\ell}$ coincide other cyclic permutation.

Monomials $\alpha_{i_1}\otimes\dotsb\otimes \alpha_{i_k}\otimes
\beta_{j_1}\otimes\dotsb\otimes\beta_{j_\ell}$ and
$\alpha_{\widetilde i_1}\otimes\dotsb\otimes \alpha_{\widetilde i_k}
\otimes \beta_{\widetilde j_1}\otimes\dotsb\otimes \beta_{\widetilde
j_\ell}$ are called {\it equivalent}, if
$\cup^k_{r=1}i_r=\cup^k_{r=1}\widetilde i_k$ and
$\cup^l_{r=1}j_r=\cup^l_{r=1}\widetilde j_r$. By
$[\alpha_{i_1}\otimes\dotsb\otimes\beta_{j_\ell}]$ denote the
equivalence class of $\alpha_{i_1}\otimes\dotsb
\otimes\beta_{j_\ell}$. A tensor series $F=\sum
c(i_1,...,i_k|j_1,...,j_\ell) a_{i_1}\otimes\dotsb \otimes
b_{j_\ell}$ generates a series $[F]=\sum
c[i_1,...,i_k|j_1,...,j_\ell][a_{i_1}\otimes\dotsb \otimes
b_{j_\ell}]$, where the sum is taken over equivalence classes of
monomials and $c[i_1,...,i_k|j_1,...,j_\ell]$ is the sum of all
coefficients $c(\widetilde i_1,...,\widetilde j_\ell)$,
corresponding to monomials from the equivalence class
$[a_{i_1}\otimes\dotsb\otimes b_{j_\ell}]$.

We say that a tensor series $F=\sum c(i_1\dotsb i_k|j_1\dotsb j_\ell)
\alpha_{i_1}\otimes\dotsb\otimes \alpha_k\otimes \beta_{j_1}\otimes
\dotsb\otimes \beta_{j_\ell}$ is a {\it Structure Series}
on a space $H=A\oplus B$ with an involution $*$ and bases
$\{\alpha_1,...,\alpha_n|\beta_1,...,\beta_m\}$, if the following
conditions hold

{\bf Axiom $1^0$.}  The coefficients $c(i_1\dotsb i_k|j_1\dotsb j_\ell)$
are invariant under all permutations of  $\{i_r\}$ and cyclic permutations of
$\{j_r\}$.

{\bf Axiom $2^0$.}  $F^*=F$.

{\bf Axiom $3^0$.}  The coefficients $c(i,j|)$ and $c(|i,j)$
generate nondegenerate matrices. By $F_a^{\alpha_i \alpha_j}$ and
$F_b^{\beta_i \beta_j}$ denote the inverse matrices of $c(i,j|)$ and
$c(|i,j)$ respectively.

{\bf Axiom $4^0$.}  $$[\sum_{p,q=1}^n\frac{\partial^3 F_a} {\partial
\alpha_i \partial \alpha_j \partial \alpha_p}\otimes F_a^{\alpha_p
\alpha_q} \frac{\partial^3 F_a}{\partial \alpha_q\partial \alpha_k
\partial \alpha_\ell}]=$$ $$[\sum_{p,q=1}^n\frac{\partial^3 F_a}{\partial
\alpha_k\partial \alpha_j
\partial \alpha_p} \otimes F_a^{\alpha_p \alpha_q}\frac{\partial^3 F_a}{\partial
\alpha_q\partial \alpha_i \partial \alpha_\ell}],$$ where $F_a$ is
the part of the tensor series $F$, consisting from all monomials
without monomials, containing some $\beta_i$.

{\bf Axiom $5^0$.}  $$[\sum_{p,q=1}^m\frac{\partial^3 F}{\partial
\beta_i \partial \beta_j
\partial \beta_p}\otimes F_b^{\beta_p \beta_q}\frac{\partial^3 F}{\partial \beta_q
\partial \beta_k \partial \beta\ell}=$$
$$\sum_{p,q=1}^m\frac{\partial^3 F}{\partial \beta_\ell\partial
\beta_i
\partial \beta_p} \otimes F_b^{\beta_p \beta_q}\frac{\partial^3 F}{\partial
\beta_q\partial \beta_j \partial \beta_k}].$$

{\bf Axiom $6^0$.}  $$[\sum\frac{\partial^2 F}{\partial \alpha_i
\partial \beta_p} \otimes F_b^{\beta_p \beta_q}\frac{\partial^3 F}{\partial
\beta_q\partial \beta_i \partial \beta_j}]= [\sum\frac{\partial^2
F}{\partial \alpha_i\partial \beta_p}\otimes F_b^{\beta_p \beta_q}
\frac{\partial^3 F}{\partial \beta_q\partial \beta_j \partial
\beta_i}].$$

{\bf Axiom $7^0$.} { $$[\sum\frac{\partial^2 F}{\partial \beta_k
\partial \alpha_p} \otimes F_a^{\alpha_p \alpha_q}\frac{\partial^3
F}{\partial \alpha_q\partial \alpha_i \partial \alpha_j}]=$$
$$[\sum\frac{\partial^2 F}{\partial \alpha_i\partial \beta_p}\otimes
F_b^{\beta_p \beta_q} \frac{\partial^3 F}{\partial \beta_q\partial
\beta_k
\partial \beta_r}\otimes F_b^{\beta_r \beta_\ell}\frac{\partial^2 F}{\partial
\beta_\ell\partial \alpha_j}].$$

We say that the conditions from axioms $1^0 - 7^0$ are
{\it Structure Equations}. They are noncommutative analogues of
associativity equations \cite{Dij, Wit2}.

\subsection{}
Any noncommutative tensor series $F=\sum
c(i_1,...,i_k|j_1,...,j_\ell) \alpha_{i_1}\otimes\dotsb\otimes
\alpha_{i_k}\otimes \beta_{j_1}\otimes\dotsb\otimes \beta_{j_\ell}$
generates a family of tensors $f^F_{r,\ell}:A^{\otimes r}\otimes
\beta^{\otimes\ell}\to\mathbb C$, where
$f^F_{r,\ell}(\alpha_{i_1}\otimes\dotsb\otimes\alpha_{i_r}
\otimes\beta_{j_1}\otimes\dotsb\otimes\beta_{j_\ell})=
c(i_1,...,i_r|j_1,...,j_\ell)$. Put $<x_{i_1},...,x_{i_r},
y_{j_1},..., y_{j_\ell}>_F=$
$f^F_{r,\ell}(x_{i_1}\otimes\dotsb\otimes x_{i_r} \otimes
y_{j_1}\otimes\dotsb\otimes y_{j_\ell})$.

{\bf Theorem 5.1.} {\sl A tensor series $F$ is a Structure Series, iff the
collection of tensors $\{<x_1,...,x_r, y_1,...,y_\ell>_F\}$
forms a System of Disk Correlation Functions.
Any System of Disk Correlation Function
is generated by a Structure Series.}

Proof: Formal calculation demonstrates equivalence the axioms of Disk
Correlation Functions and the axioms of Structure Series with the
same numbers. $\square$

\section {Noncommutative Frobenius manifolds}

\subsection{}
According to \cite{Dub}, any solution WDVV equation generates
a special deformation of Frobenius algebra. In this
section we associate to any solution of Structure
Equation some deformation of Extended Frobenius algebra.

\begin{definition} Extended Frobenius algebra
$\mathcal H=\{H=A\dotplus B, (x,y), x\mapsto x^*\}$
is a finite dimensional associative algebra $H$ over $\mathbb K$
endowed with

a decomposition $H=A\dotplus B$ of $H$ into a direct sum of
vector spaces;

an invariant symmetric bilinear form $(x,y): H\otimes H\to\mathbb K$
that is $(x, y)= (y, x)$ and $(xy, z)=(x, yz)$;

an involutive anti-automorphism $H\to H$, denoted by $x\mapsto x^*$;
\\ such that the following axioms hold:

$1^\circ$ $A$ is a subalgebra belonging to the centre of algebra $H$;

$2^\circ$ $B$ is a two-sided ideal of $H$ (typically noncommutative);

$3^\circ$ restrictions $(x,y)|_A$ and $(x,y)|_B$ are
nondegenerate scalar products on algebras $A$ and $B$ resp.

$4^\circ$ an involutive anti-automorphism preserves the
decomposition $H=A\dotplus B$ and the form $(x,y)$ on $H$, i.e.
$A^*=A$, $B^*=B$, $(x^*,y^*)=(x,y)$.
\end{definition}

A full description of semisimple Extended
Frobenius algebras over $\mathbb C$ follows from \cite{Al-Nat} $\S 2$.

The structure tensors
$F_{\alpha',\alpha''}=(\alpha',\alpha''), F_{\beta',\beta''}=
(\beta',\beta''), R_{\alpha\beta}=(\alpha,\beta),
S_{\alpha',\alpha'',\alpha'''}=(\alpha'\alpha'',\alpha'''),
T_{\beta'\beta''\beta'''}=(\beta'\beta'',\beta''')$,
$R_{\alpha\beta'\beta''}=(\alpha\beta',\beta'')$,
$I_{\alpha'\alpha''}=((\alpha')^*,\alpha'')$,
$I_{\beta'\beta''}=((\beta')^*,\beta'')$,
uniquely describe an Extended Frobenius algebra \cite{Al-Nat}.
Here we denote by $\alpha,\alpha',\alpha''$ and $\beta,\beta',\beta''$
the elements of the bases $\{\alpha_1,...,\alpha_n\}\subset A$ and
$\{\beta_1,...,\beta_m\}\subset B$ respectively.

\subsection{}
Later we assume that the field $\mathbb K$ is the field of real or
complex numbers. Associate a deformation of Extended Frobenius
algebras to a Structure Series on a space $H=A + B$
with involution $*:H\to H$ and a basis
$\{\alpha_1,...,\beta_\ell\}\subset H$. First consider
the coordinates on $H$ that associate the element
$z=s^i\alpha_i+t^j\beta_j$ to collections
$(s^1,...,s^n|t^1,...,t^m)\in \mathbb K^{n+m}$.
The correspondences $\alpha_i\mapsto s^i$, $\beta_i\mapsto t^i$
and (tensor multiplication) $\mapsto$(number multiplication)
map tensor series $F(\alpha_1,..., \beta_m)$ to formal number series
$\hat F(s^1,...,t^m)$. By definition, put
$F_{\alpha_i\alpha_j}=c(i,j|)$, $F_{\beta_i\beta_j}=c(|i,j)$,
$I_{\alpha_i\alpha_j}=F_{\alpha_i\alpha_k^*}$,
$I_{\beta_i\beta_j}=F_{\beta_i\beta_k^*}$.
Let $\{F^{\alpha_i\alpha_j}\}$ and $\{F^{\beta_i\beta_j}\}$
be the inverse matrix for $\{F_{\alpha_i\alpha_j}\}$ and
$\{F_{\beta_i\beta_j}\}$ respectively. Put
$$R^F_{\alpha_i\beta_j} =\frac{\partial^2 F}
{\partial \alpha_i\partial \beta_j}, \
S^F_{\alpha_i\alpha_j\alpha_k} =\frac{\partial^3 F} {\partial
\alpha_i
\partial \alpha_j \partial \alpha_k},$$
$$T^F_{\beta_i\beta_j\beta_k} =
\frac{\partial^3  F}{\partial \beta_i\partial \beta_j\partial
\beta_k},$$
$$R^F_{\alpha_i\beta_j\beta_k}=
\frac{\partial^2  F}{\partial \alpha_i\partial \beta_p} F^{\beta_p
\beta_q} \frac{\partial F}{\partial \beta_q\partial \beta_j\partial
\beta_k}.$$

{\bf Theorem 6.1.} {\sl Let $F$ be a Structure Series. Then at
convergence points the tensors $F_{\alpha_i\alpha_j}$,
$F_{\beta_i\beta_j}$, $I_{\alpha_i\alpha_j}$, $I_{\beta_i\beta_j}$,
$\hat R_{\alpha_i\beta_j} =\hat R_{\alpha_i\beta_j}^F(s^1,...,t^m)$,
$\hat S_{\alpha_i\alpha_j\alpha_k} =\hat
S^F_{\alpha_i\alpha_j\alpha_k} (s^1,...,t^m)$, $\hat
T_{\beta_i\beta_j\beta_k} =\hat T^F_{\beta_i\beta_j\beta_k}
(s^1,...,t^m)$, $\hat R_{\alpha^i\beta^j\beta^k} =\hat
R^F_{\alpha^i\beta^j\beta^k} (s^1,...,t^m)$ define an Extended
Frobenius algebra.}

Proof: We shall raise indices of tensors, using the tensors
$F^{\alpha_i\alpha_j}$ and $F^{\beta_i\beta_j}$. In the case of
non-symmetric tensors we always raise the last index. By definition,
put $\hat S_{\alpha_i\alpha_j\alpha_k\alpha_\ell}=\sum_r \hat
S^{\alpha_r}_{\alpha_i\alpha_j} \hat
S_{\alpha_r\alpha_k\alpha_\ell}$ $\hat
T_{\beta_i\beta_j\beta_k\beta_\ell}=\sum_r \hat T^{\beta_r}
_{\beta_i\beta_j} \hat T_{\beta_r\beta_k\beta_\ell}$. It follows
from \cite{Al-Nat} that the above theorem is equivalent to the
following conditions:

1) Matrices  $F_{\alpha_i\alpha_j}$ and $F_{\beta_i\beta_j}$ are nondegenerate;

2) Tensors  $\hat S_{\alpha_i\alpha_j\alpha_k}$ and
$\hat S_{\alpha_i\alpha_j\alpha_k\alpha_\ell}$ are symmetric with respect to
all permutations;

3) Tensors  $\hat T_{\beta_i\beta_j\beta_k}$ and
$\hat T_{\beta_i\beta_j\beta_k\beta_\ell}$ are symmetric with respect to
cyclic permutations;

4) $\hat R_{\alpha,\beta_1,\beta_2}=
\hat R_\alpha^{\beta'} \hat T_{\beta',\beta_1,\beta_2}$;

5) $\hat R_{\beta}^{\alpha'}\hat S_{\alpha',\alpha_1,\alpha_2}=
        \hat R_{\alpha_1}^{\beta'}\hat R_{\alpha_2}^{\beta''}
        \hat T_{\beta',\beta'',\beta}$;

6) $\hat R_{\alpha,\beta_1,\beta_2}=\hat R_{\alpha,\beta_2,\beta_1}$;

7) $\hat I_{\alpha_1}^{\alpha'}\hat I_{\alpha',\alpha_2}
=F_{\alpha_1,\alpha_2}$, $\hat I_{\beta_1}^{\beta'}\hat
I_{\beta',\beta_2}=F_{\beta_1,\beta_2}$;

8) $\hat I_{\alpha_1,\alpha_2}=\hat I_{\alpha_2,\alpha_1}$, $\hat
I_{\beta_1,\beta_2}=\hat I_{\beta_2,\beta_1}$, $\hat
I_{\alpha}^{\alpha'}\hat R_{\alpha',\beta}= \hat
I_{\beta}^{\beta'}\hat R_{\alpha,\beta'}$;

9) $\hat I_{\alpha_1}^{\alpha'}\hat I_{\alpha_2}^{\alpha''} \hat
I_{\alpha_3}^{\alpha'''} \hat S_{\alpha''',\alpha'',\alpha'} =\hat
S_{\alpha_1,\alpha_2,\alpha_3}$, $\hat I_{\beta_1}^{\beta'}\hat
I_{\beta_2}^{\beta''} \hat I_{\beta_3}^{\beta'''} \hat
T_{\beta''',\beta'',\beta'}=\hat T_{\beta_1,\beta_2,\beta_3}$.

All these conditions directly follow from our definition and
the axioms of Structure Series. $\square$

Thus, Structure Series generate an Extended Frobenius algebra in
the points of $H$, where the series from theorem 6.1
converge. This defines a deformation of Extended Frobenius
algebra i.e. some analog of Frobenius manifold \cite{Dub}
for (noncommutative) Extended Frobenius algebras.

\section{Examples}

\subsection{}
For $\text{dim}\ B=0$ theorems 4.1 and 5.1 get over into theorem
III, 4.3 from \cite{Man} that claim that Cohomological Field Theory
is equivalent to Formal Frobenius manifolds.

Noncommutative Frobenius manifolds with $\text{dim} A=0,
\text{dim} B=1$ are Frobenius manifolds of rang 1. They
are described by formal series $F(\beta)$ \cite{Man}.

For $\text{dim} A=1, \text{dim} B=1$. Put
$<\alpha^n, \beta^m>=<\alpha,...,\alpha, \beta,...,\beta>$.
Axiom $7^0$ ($\S 5$) demonstrate that all correlators
are determined by correlators
$<\alpha^n>, <\beta^m>$ and $<\alpha, \beta>$.

Suppose for example that
$<\alpha,\alpha>=<\alpha,\alpha,\alpha>=<\beta,\beta>
=<\beta,\beta,\beta>=1$ and $<\alpha^m>=<\beta^m>=0$ for
$m>3$. Then according to Axiom $7^0$,
$<\alpha^3><\alpha,\beta>=<\beta^3><\alpha,\beta><\alpha,\beta>$
and thus $<\alpha,\beta>$ is equal to 0 or 1.
Moreover, Axiom $7^0$ gives
$<\alpha^3><\alpha\beta^2>=2<\alpha,\beta><\beta^4><\alpha,\beta>
+2<\alpha,\beta><\beta^3><\alpha,\beta^2>$ and thus
$<\alpha,\beta^2>=0$. Analogously we prove that
$<\alpha,\beta^m>=0$ for $m>1$.
On the other hand Axiom $7^0$ gives
$<\alpha^3><\alpha^2,\beta>=<\alpha,\beta><\alpha\beta><\alpha,\beta^3>
+2<\alpha^2,\beta><\alpha,\beta><\beta^3>$ and thus
$<\alpha^2,\beta>=0$. Analogously we prove that
$<\alpha^n,\beta^m>=0$ for $nm>1$.

Thus, if the restriction of Structure Series $F$ on $A$ and $B$
is $\frac 12\alpha^{\otimes 2}+\alpha^{\otimes 3}$
and $\frac 12\beta^{\otimes 2}+\beta^{\otimes 3}$
then $F=\frac 12 a^{\otimes 2}+ a^{\otimes 3}+
\frac 12 b^{\otimes 2}+b^{\otimes 3}+
\xi\alpha\otimes\beta$, where $\xi\in\{0, 1\}$.

\subsection{}
In analogy with \cite{Nat-Tur} any Extended Frobenius algebra
$\mathcal H=\{H=A\dot + B, (. , .), * \}$ with unite 1
generates some System of Disk Correlation Functions by
the rule $<\alpha^1,...,\alpha^s, \beta^1,...,\beta^t>=
(\alpha^1\dotsb \alpha^s\beta^1\dotsb \beta^t, 1)$.

\subsection{}
According to \cite{Br}, a noncommutative Frobenius manifold in our
sense is generated by a topological Landay-Ginzburg theory of
$B$-type with $2D$-brane conditions.

\bigskip

S.M.~Natanzon

Moscow State University

Institute Theoretical and Experimental Physics

Independent University of Moscow

natanzon@mccme.ru

\end{document}